\documentclass{article}
\usepackage[utf8]{inputenc}
\usepackage[margin=1in]{geometry}
\usepackage{slashed,graphicx}
\usepackage{amsthm}
\usepackage{amsmath}
\usepackage{float}
\usepackage{amssymb}
\usepackage[caption=false]{subfig}
\usepackage{hyperref}
\title{Placing of the recently observed  bottom strange state $B_{sJ}(6063)$ and $B_{sJ}(6114)$ in bottom spectra}
\author{Ritu Garg, Pallavi Gupta, A. Upadhyay
 \\\small{\it School of Physics and Material Science},\\\small{\it Thapar University,
Patiala, Punjab-147004},\\\small{\it Mehr Chand Mahajan DAV college for Women, Chandigarh}
\\\small{E-mail: ritugarg039@gmail.com,
alka.iisc@gmail.com}}

\begin{document}

\maketitle

\section{Abstract}
We have employed HQET to give the spin-parity quantum numbers for recently observed bottom strange states $B_{sJ}(6063)$ and $B_{sJ}(6114)$ by LHCb collaborations. By exploring flavour independent parameters $ \Delta_{F}^{(c)} =\Delta_{F}^{(b)}$ and $ \lambda_{F}^{(c)} = \lambda_{F}^{(b)}$, we calculated masses of experimentally missing bottom strange meson states $2S, 1P, 1D$. We have also analyzed these bottom strange masses by taking ${1/m_Q}$ corrections which lead modifications of parameter terms as $ \Delta_{F}^{(b)} =\Delta_{F}^{(c)} + \delta\Delta_F$ and $ \lambda_{F}^{(b)} = \lambda_{F}^{(c)}\delta\lambda_F$. Further, we have analyzed their two-body decays, couplings, and branching ratios via the emission of light pseudoscalar mesons. Based on predicted masses and decay widths, we tentatively identified the states $B_{sJ}(6063)$ as $2^3S_1$ and $B_{sJ}(6114)$ as $1^3D_1$. Our predictions provide crucial information for future experimental studies.
\section{Introduction}
During the last decades, different experimental facilities like LHCb, BABAR, BESIII, FOCUS, SLAC, etc., have been on a discovery spree for stimulating the spectrum of heavy-light mesons. Based on flavour of the heavy quark, heavy-light mesons can be cataloged into the charm and bottom mesons. In the charm meson sector, observations of some ground and excited states like $D_{0}(2550)$, $D_{1}^{*}(2600)$, $D_{2}(2740)$, $D_{3}^{*}(2750)$, $D^{0}(3000)$, $D_{J}^{*}(3000)$ and strange states $D_{s1}(2860)$, $D_{sJ}(3040)$, $D_{s0}(2590)$ \cite{1,2,3,4,5,6,7} have not only broaden the spectra but also help us in exploring their properties through decay studies. However, experimental growth toward establishing the bottom sector is still lacking. Only ground states $B^{0,\pm}(5279)$, $B^{*}(5324)$, $B_{s}(5366)$, $B_{s}^{*}(5415)$ and some low lying states $B_{1}(5721)$, $B_{J}^{*}(5732)$, $B_{2}^{*}(5747)$, $B_{s1}(5830)$, $B_{s2}^{*}(5840)$, $B_{sJ}^{*}(5850)$, $B_{J}(5840)$, $B_{J}(5970)$ are observed experimentally and listed in PDG \cite{8}. But apart from these states, the whole bottom meson spectra are unknown. To fill this gap, experimentalists, and theoretical models are trying to predict new states that could fill this gap. In this process, recently, LHCb collaborations discovered two new states $B_{sJ}(6063)$ and $B_{sJ}(6114)$ in $B^{+}K^{-}$ mass spectrum \cite{9}. The measured masses and decay widths are given below:\\\\
$M(B_{sJ}(6063))= 6063.5\pm1.2(stat)\pm0.8(syst) MeV$\\
$\Gamma(B_{sJ}(6063)) = 26\pm4\pm4$ $MeV/c^2$\\\\
$M(B_{sJ}(6114))= 6114\pm3(stat)\pm5(syst) MeV$\\
$\Gamma(B_{sJ}(6114)) = 66\pm18\pm21$ $MeV/c^2$\\\\

The successes of the observations of these radially excited states by LHCb have demonstrated that more excited bottom meson states will be discovered in future LHC experiments.
 Exploring higher excited B mesons is not continuing. LEP (Large electron-positron collider) in 1999 reported an orbitally excited bottom meson state in the hadronic Z decays process \cite{10}. This bottom meson's measured mass, and decay width is $5937\pm21\pm4 MeV$ and $50\pm22\pm5 MeV$, respectively. However, this state was never reconfirmed by other experimental facilities. After many years, CDF collaboration in 2013 observed new resonance $B(5970)$ in decay modes $B^0\pi^+$ and $B^+\pi^-$ simultaneously \cite{11}. Two years after LHCb collaboration in 2015, four resonances $B_J(5840)^{0,+}$ and $B_J(5960)^{0,+}$ were announced \cite{12}. Despite these observations of mesons, the spectrum of excited bottom mesons is not much explored. In strange bottom mesons family, only some states $B_{s}(5366)$, $B^{*}_{s}(5415)$, $B_{s1}(5830)$, $B^{*}_{s2}(5840)$ are well established and collected in PDG.\cite{8}. Among them, states $B_{s}(5366)$, $B^{*}_{s}(5415)$ are classified as $1S$ states and states $B_{s1}(5830)$, $B^{*}_{s2}(5840)$ are assigned as $1P$ ($1^+, 2^+$) states. This shows experimentalists continue trying to establish the bottom strange meson spectrum.\\
In theory, various theoretical studies have performed different analyses for higher excited bottom non-strange and bottom strange meson states \cite{13,14,15,16,17,18,19,20,21,22,23,24,25,26,27,28,29,30,31,32,33}. With the help of theoretical models, states $B^{0,\pm}(5279)$, $B^{*}(5324)$, $B_{s}(5366)$, $B_{s}^{*}(5415)$ are assigned as $1S$ state which very well matches with experimental data. Further the states $B_{1}(5721)$, $B_{2}^{*}(5747)$ are also well established experiemntally and classified as $1P(1^+, 2^+)$ respectively. But theoretically, state $B_{1}(5721)$ is still a disputed candidate because some of the theoretical work with heavy meson effective theory favors it as $1P(1^+)$ state \cite{25,32}, while other work using relativistic quark model and non-relativistic quark model explained this state as a mixture of $^3P_1$ and $^1P_1$ states \cite{14,16,33}. The $J^P$ of state $B_J(5840)$ is still ambiguous as different models suggest different $J^P$'s state for it. Authors in \cite{16,17} explained $B_J(5840)$ with the quark model and suggested the assignment as $2^1S_0$ while G. L. Yu and Z. G. Wang with $^3 P_0$ decay model analysis favors the assignments of state $B_J(5840)$ as $2^3S_1$ \cite{20}. But Heavy quark effective theory explained resonances $B_J(5840)$ as $1^3D_1$ state \cite{28}. The state $B_J(5960)^{0,+}$ is assigned as $2^3S_1$ or $1^3D_3$ state or $1^3D_1$ state with different theoretical models \cite{14,15,16,22,23,24,34}. But its $J^P$ value is still a question mark in PDG, which only mention its mass and decay width. We have discussed here a brief literature on these non-strange bottom states. The assignments of these states ($B_{1}(5721)$, $B_{2}^{*}(5747)$, $B_J(5970)$) are also suggested in our previous work \cite{28}. In the case of the strange bottom sector, only a few states have been observed, out of which $B_{s1}(5830)$, $B_{s2}(5840)$ states are well observed by CDF \cite{11,35}, D0 \cite{36}, and LHCb \cite{37} collaborations and are identified as $1P(1^+, 2^+)$ respectively. But, there is ambiguity for recently observed strange bottom meson states $B_{sJ}(6063)$ and $B_{sJ}(6114)$. The states $B_{sJ}(6063)$ and $B_{sJ}(6114)$ in non-relativistic quark potential model identified as $1^3D_1$ and $1^3D_3$ states respectively \cite{38,39}. While authors in Ref. \cite{40} assign these states as $1^3D_1$ and $2^3S_1$ states, respectively. Theoretical analysis for these newly observed states is thus limited in the literature, indicating it needs more attention. As a continuation of previous work \cite{28}, we analyzed observed strange bottom mesons states $B_{sJ}(6063)$ and $B_{sJ}(6114)$ and give their $J^P$ values within this framework. \\
The paper is arranged as follows: Section II briefly describes the model “Heavy Quark Effective Theory.” Section III represents the numerical analysis of $B_{sJ}(6063)$ and $B_{sJ}(6114)$ based on predicted masses and decay widths. Section IV presents our final conclusion. 

\section{Framework}
We apply heavy quark effective theory to assign spin-parity quantum numbers for recently observed heavy-light strange bottom meson states. This theory is simple and powerful, which provides precise calculations of masses, and decay behavior of heavy-light bottom mesons \cite{41}. HQET assumes an infinite mass of heavy quarks (Q = c, b) and most of the momentum of the bottom meson is carried by heavy quark. In this heavy quark limit ($m_Q\rightarrow \infty$), spin of the heavy quark $s_Q$ decouples from the light d.o.f.(degree of freedom), which incorporates the light antiquark and the gluons. The total angular momentum of light d.o.f is $s_{l} = s_{q} + l$, where $s_{q}$ = 1/2, the spin of light quark and $l$ is the total orbital momentum of light quarks. In heavy quark limit, mesons are categorized in doublets based on the total angular momentum of light quarks. For $l = 0$, $s_{l} = 1/2$ associate with spin of heavy quark $s_{Q}$ = 1/2 and resulted with doublet $(0^{-},1^{-})$. This doublet is denoted by $(P, P^{*})$. While $l = 1$ forming two doublets are given by $(P^{*}_{0}, P_{1}^{'})$ and $(P_{1}, P_{2}^{*})$ with $J_{s_{l}}^{P} = (0^{+}, 1^{+})_{1/2}$ and $J_{s_{l}}^{P} = (1^{+}, 2^{+})_{3/2}$ respectively. For $l = 2$, two doublets are expressed by $(P^{*}_{1}, P_{2})$ and $(P_{2}^{'},P_{3}^{*})$ with $J_{s_{l}}^{P} = (1^{-}, 2^{-})_{3/2}$ and $J_{s_{l}}^{P} = (2^{-}, 3^{-})_{5/2}$ respectively. These doublets are introduced in terms of super effective fields $H_{a}, S_{a}, T_{a}, X^{\mu}_{a}, Y^{\mu\nu}_{a}$ and described for fields are given below \cite{25,42}:
\begin{gather}
\label{eq:lagrangian1}
 H_{a}=\frac{1+\slashed
v}{2}\{P^{*}_{a\mu}\gamma^{\mu}-P_{a}\gamma_{5}\}\\
S_{a} =\frac{1+\slashed v}{2}[{P^{'\mu}_{1a}\gamma_{\mu}\gamma_{5}}-{P_{0a}^{*}}]\\
T^{\mu}_{a}=\frac{1+\slashed v}{2}
\{P^{*\mu\nu}_{2a}\gamma_{\nu}-P_{1a\nu}\sqrt{\frac{3}{2}}\gamma_{5}
[g^{\mu\nu}-\frac{\gamma^{\nu}(\gamma^{\mu}-\upsilon^{\mu})}{3}]\}
\end{gather}
\begin{gather}
X^{\mu}_{a}=\frac{1+\slashed
v}{2}\{P^{\mu\nu}_{2a}\gamma_{5}\gamma_{\nu}-P^{*}_{1a\nu}\sqrt{\frac{3}{2}}[g^{\mu\nu}-\frac{\gamma_{\nu}(\gamma^{\mu}+v^{\mu})}{3}]\}
\end{gather}

\begin{multline}
 Y^{\mu\nu}_{a}=\frac{1+\slashed
v}{2}\{P^{*\mu\nu\sigma}_{3a}\gamma_{\sigma}-P^{'\alpha\beta}_{2a}\sqrt{\frac{5}{3}}\gamma_{5}[g^{\mu}_{\alpha}g^{\nu}_{\beta}
-\frac{g^{\nu}_{\beta}\gamma_{\alpha}(\gamma^{\mu}-v^{\mu})}{5}-\frac{g^{\mu}_{\alpha}\gamma_{\beta}(\gamma^{\nu}-v^{\nu})}{5}]\}
\end{multline}
The field $H_a$ describes doublets of $S$-wave for $J^P = (0^-, 1^-)$. The fields $S_a$ and $T_a$ represent doublets of $P$-wave for $J^P = (0^+, 1^+)$ and $(1^+, 2^+)$ respectively. $D$-wave doublets for $J^P = (1^-, 2^-)$ and $(2^-, 3^-)$ belong to fields $X^{\mu}_{a}$ and $Y^{\mu\nu}_{a}$ respectively. $a$ in above expressions is light quark ($u, d, s$) flavour index. $v$ is heavy quark velocity, unchanged in strong interactions. The approximate chiral symmetry $SU(3)_L\times SU(3)_R$ is incorporated with fields of pseudoscalar mesons $\pi$, K, and $\eta$ that are the lightest strongly interacting bosons. They are considered as approximate Goldstone bosons of this chiral symmetry and can be expressed by the matrix field $\xi=e^{\frac{i\mathcal{M}}{f_{\pi}}}$ and $\Sigma=\xi^{2}$, where $\mathcal{M}$ is given by
\begin{center}
\begin{equation}
\mathcal{M}  = \begin{pmatrix}
\frac{1}{\sqrt{2}}\pi^{0}+\frac{1}{\sqrt{6}}\eta & \pi^{+} & K^{+}\\
\pi^{-} & -\frac{1}{\sqrt{2}}\pi^{0}+\frac{1}{\sqrt{6}}\eta &
K^{0}\\
K^{-} & \overline{K}^{0} & -\sqrt{\frac{2}{3}}\eta
\end{pmatrix}
\end{equation}
\end{center}
 $f_{\pi}$ is pion decay constant, and its value is taken 130 $MeV$. Fields of heavy meson doublets is given in eqn.(1-5) interact with pseudoscalar goldstone bosons through covariant derivative $D_{\mu ab}= -\delta_{ab}\partial_{\mu}+\mathcal{V}_{\mu ab} =  -\delta_{ab}\partial_{\mu}+\frac{1}{2}(\xi^{+}\partial_{\mu}\xi+\xi\partial_{\mu}\xi^{+})_{ab}$ and axial vector field $A_{\mu ab}=\frac{i}{2}(\xi\partial_{\mu}\xi^{\dag}-\xi^{\dag}\partial_{\mu}\xi)_{ab}$. By comprising all meson doublet fields and goldstone fields, effective lagrangian is written as:

\begin{multline}
\mathcal{L} = iTr[\overline{H}_{b}v^{\mu}D_{\mu ba}H_{a}]+ \frac{f_\pi^{2}}{8}Tr[\partial^{\mu}\Sigma\partial_{\mu}\Sigma^{+}] + Tr[\overline{S_{b}}(iv^{\mu}D_{\mu ba} - \delta_{ba}\Delta_{S})S_{a}]+Tr[\overline{T_{b}^{\alpha}}(iv^{\mu}D_{\mu ba}- \delta_{ba}\Delta_{T})T_{a \alpha}\\+ Tr[\overline{X_{b}^{\alpha}}(iv^{\mu}D_{\mu ba}- \delta_{ba}\Delta_{X})X_{a \alpha}+ Tr[\overline{Y_{b}^{\alpha\beta}}(iv^{\mu}D_{\mu ba}- \delta_{ba}\Delta_{Y})Y_{a\alpha\beta}
\end{multline}
The mass parameter $\Delta_{F}$ in equation (7) gives the mass difference between excited mass doublets ($F$) and ground mass doublet ($H$) in the form of spin average masses of these doublets with the same principle quantum number ( n ). The mass parameters are described by:
\begin{align}
             \Delta_{F}=\overline{M_{F}}&- \overline{M_{H}},~~ F= S,T,X,Y\\
\text{where, }~~~~~~~~~~~
           \overline{M_{H}}&=(3m^{Q}_{P^*}+m^{Q}_{P})/4\\
         \overline{M_{S}}&=(3m^{Q}_{P_1^{'}}+m^{Q}_{P_0^*})/4\\
        \overline{M_{T}}&=(5m^{Q}_{P_2^*}+3m^{Q}_{P_1})/8\\
          \overline{M_{X}}&=(5m^{Q}_{P_2}+3m^{Q}_{P_1^{*}})/8\\
           \overline{M_{Y}}&=(7m^{Q}_{P_3^*}+5m^{Q}_{P_2^{'}})/12
         \end{align}
           The $1/m_{Q}$ corrections to heavy quark limit are given by symmetry-breaking terms. The corrections are form of: 
           \begin{multline}
          \mathcal{L}_{1/m_{Q}} = \frac{1}{2m_{Q}}[\lambda_{H} Tr(\overline H_{a}\sigma^{\mu\nu}{H_{a}}\sigma_{\mu\nu}) - \lambda_{S}Tr(\overline S_{a}\sigma^{\mu\nu} S_{a}\sigma_{\mu\nu})+\lambda_{T}Tr(\overline T_{a}^{\alpha}\sigma^{\mu\nu}{T_{a}^{\alpha}}\sigma_{\mu\nu})-\lambda_{X}Tr(\overline X_{a}^{\alpha}\sigma^{\mu\nu}{X_{a}^{\alpha}}\sigma_{\mu\nu})\\ + \lambda_{Y}Tr(\overline Y_{a}^{\alpha\beta}\sigma^{\mu\nu}{Y_{a}^{\alpha\beta}}\sigma_{\mu\nu}) 
          \end{multline} 
          Here parameters $\lambda_{H}$, $\lambda_{S}$, $\lambda_{T}$, $\lambda_{X}$, $\lambda_{Y}$ are analogous with hyperfine splittings and expressed in Eq.(15 - 19). These mass terms in lagrangian give only first order in $1/m_{Q}$ terms, but higher order terms may also be present otherwise. We are limited to the first order corrections in $1/m_{Q}$.
 \begin{align}
     \lambda_{H} = \frac{1}{8}(M^{2}_{P^{*}} - M^{2}_{P}) \\
     \lambda_{S} = \frac{1}{8}({M^{2}_{P_1^{'}}} - {M^{2}_{P_0^*}})\\
     \lambda_{T}=\frac{3}{8}({M^{2}_{P_2^*}}-{M^{2}_{P_1}})\\
     \lambda_{X}=\frac{3}{8}({M^{2}_{P_2}}-{M^{2}_{P_1^*}})\\
      \lambda_{Y}=\frac{3}{8}({M^{2}_{P_3}}-{M^{2}_{P_2^{'*}}})
 \end{align}

   In HQET, at the scale of 1 $GeV$, flavour symmetry spontaneously arises for b (bottom quark) and c (charm quark) and hence beauty of flavour symmetry implies to
        \begin{align}
           \label{1eu_eqn}
           \Delta_{F}^{(c)} =\Delta_{F}^{(b)}\\
   \lambda_{F}^{(c)} = \lambda_{F}^{(b)}
   \label{2eu_eqn}
\end{align}
This symmetry is broken by the higher order terms in the HQET Lagrangian involving terms of factor $1/m_{Q}$ and the parameters $\Delta_{F}$ and $\lambda_{F}$ are modified by extra terms $\delta\Delta_{F}$ and $\delta\lambda_{F}$.
The hyperfine splitting term $\lambda_{F}$, which originates from the chromomagnetic interaction, is dominated by the QCD corrections and the $1/m_{Q}$ effect is neglected. QCD corrections change the $\lambda_{F}$ relation to\\
$\lambda_{F}^{(b)} = \lambda_{F}^{(c)}{{(\frac{\alpha_s(m_b)}{\alpha_s(m_c)})}^{9/25}}$.
\\The difference of the spin averaged masses at $1/m_{Q}$ order modifies the parameter $\Delta_{F}$ by  $\delta\Delta_{F}$ given by
$$\Delta_{F}^{(b)}
 = \Delta_{F}^{(c)}
 + \delta\Delta_{F}$$ ,

The decays $F\rightarrow H + M$ (F = $H, S, T, X, Y$
and $M$ presents a light pseudoscalar meson) can be described by effective Lagrangians explained in terms of the fields introduced in (9-14) that are valid at leading order in the heavy quark mass and light meson momentum expansion:
\begin{gather}
\label{eq:lagrangian17}
L_{HH}=g_{HH}Tr\{\overline{H}_{a}
H_{b}\gamma_{\mu}\gamma_{5}A^{\mu}_{ba}\}\\
L_{SH}=g_{SH}Tr\{\overline{H}_{a}S_{b}\gamma_{\mu}\gamma_{5}A^{\mu}_{ba}\}+h.c.\\
L_{TH}=\frac{g_{TH}}{\Lambda}Tr\{\overline{H}_{a}T^{\mu}_{b}(iD_{\mu}\slashed
A + i\slashed D A_{\mu})_{ba}\gamma_{5}\}+h.c.\\
L_{XH}=\frac{g_{XH}}{\Lambda}Tr\{\overline{H}_{a}X^{\mu}_{b}(iD_{\mu}\slashed
A + i\slashed D A_{\mu})_{ba}\gamma_{5}\}+h.c.
\end{gather}
\begin{multline}
L_{YH}=\frac{1}{\Lambda^{2}}Tr\{\overline{H}_{a}Y^{\mu\nu}_{b}[k^{Y}_{1}\{D_{\mu}
,D_{\nu}\}A_{\lambda}+k^{Y}_{2}(D_{\mu}D_{\lambda}A_{\nu}
+D_{\nu}D_{\lambda}A_{\mu})]_{ba}\gamma^{\lambda}\gamma_{5}\}+h.c.
\end{multline}

In these equations $D_{\mu} =
\partial_{\mu}+V_{\mu}$,  $\{D_{\mu},D_{\nu}\}
= D_{\mu}D_{\nu}+D_{\nu}D_{\mu}$ and $\{D_{\mu} ,D_{\nu}D_{\rho}\} =
D_{\mu}D_{\nu}D_{\rho}+D_{\mu}D_{\rho}D_{\nu}+D_{\nu}D_{\mu}D_{\rho}+D_{\nu}D_{\rho}D_{\mu}+D_{\rho}D_{\mu}
D_{\nu}+D_{\rho}D_{\nu}D_{\mu}$. $\Lambda$ is the chiral symmetry
breaking scale taken as 1 $GeV$. $g_{HH}$, $g_{SH}$, $g_{TH}$, $g_{YH}
= k^{Y}_{1}+k^{Y}_{2}$ are the
strong coupling constants involved.  Using the lagrangians
$L_{HH}, L_{SH}, L_{TH}, L_{YH}$, the two body strong decays of $Q\overline{q}$ heavy-light bottom mesons are given as \cite{43,44,45,46}\\\\
\\$(0^{-},1^{-}) \rightarrow (0^{-},1^{-}) + M$
\begin{gather}
\label{eq:lagrangian} \Gamma(1^{-} \rightarrow 1^{-})=
C_{M}\frac{g_{HH}^{2}M_{f}p_{M}^{3}}{3\pi f_{\pi}^{2}M_{i}}\\
\Gamma(1^{-} \rightarrow 0^{-})=
C_{M}\frac{g_{HH}^{2}M_{f}p_{M}^{3}}{6\pi f_{\pi}^{2}M_{i}}\\
\Gamma(0^{-} \rightarrow 1^{-})=
C_{M}\frac{g_{HH}^{2}M_{f}p_{M}^{3}}{2\pi f_{\pi}^{2}M_{i}}
\end{gather}

 $(0^{+},1^{+}) \rightarrow (0^{-},1^{-}) + M$
\begin{gather}
\label{eq:lagrangian} \Gamma(1^{+} \rightarrow 1^{-})=
C_{M}\frac{g_{SH}^{2}M_{f}(p^{2}_{M}+m^{2}_{M})p_{M}}{2\pi f_{\pi}^{2}M_{i}}\\
\Gamma(0^{+} \rightarrow 0^{-})=
C_{M}\frac{g_{SH}^{2}M_{f}(p^{2}_{M}+m^{2}_{M})p_{M}}{2\pi
f_{\pi}^{2}M_{i}}
\end{gather}

 $(1^{+},2^{+}) \rightarrow (0^{-},1^{-}) + M$
\begin{gather}
\label{eq:lagrangian} \Gamma(2^{+} \rightarrow 1^{-})=
C_{M}\frac{2g_{TH}^{2}M_{f}p_{M}^{5}}{5\pi f_{\pi}^{2}\Lambda^{2}M_{i}}\\
\Gamma(2^{+} \rightarrow 0^{-})=
C_{M}\frac{4g_{TH}^{2}M_{f}p_{M}^{5}}{15\pi f_{\pi}^{2}\Lambda^{2}M_{i}}\\
\Gamma(1^{+} \rightarrow 1^{-})=
C_{M}\frac{2g_{TH}^{2}M_{f}p_{M}^{5}}{3\pi
f_{\pi}^{2}\Lambda^{2}M_{i}}
\end{gather}

$(1^{-},2^{-}) \rightarrow (0^{-},1^{-}) + M$
\begin{gather}
\label{eq:lagrangian} \Gamma(1^{-} \rightarrow 0^{-})=
C_{M}\frac{4g_{XH}^{2}}{9\pi f_{\pi}^{2}\Lambda^{2}}
\frac{M_{f}}{M_{i}}[p_{M}^{3}(m_{M}^{2}+p_{M}^{2})]\\
\Gamma(1^{-} \rightarrow 1^{-})= C_{M}\frac{2g_{XH}^{2}}{9\pi
f_{\pi}^{2}\Lambda^{2}}
\frac{M_{f}}{M_{i}}[p_{M}^{3}(m_{M}^{2}+p_{M}^{2})]\\
\Gamma(2^{-} \rightarrow 1^{-})= C_{M}\frac{2g_{XH}^{2}}{3\pi
f_{\pi}^{2}\Lambda^{2}}
\frac{M_{f}}{M_{i}}[p_{M}^{3}(m_{M}^{2}+p_{M}^{2})]
\end{gather}

$(2^{-},3^{-}) \rightarrow (0^{-},1^{-}) + M$

\begin{gather}
\label{eq:lagrangian} \Gamma(2^{-} \rightarrow 1^{-})=
C_{M}\frac{4g_{YH}^{2}}{15\pi f_{\pi}^{2}\Lambda^{4}}
\frac{M_{f}}{M_{i}}[p_{M}^{7}]\\
\Gamma(3^{-} \rightarrow 0^{-})= C_{M}\frac{4g_{YH}^{2}}{35\pi
f_{\pi}^{2}\Lambda^{4}} \frac{M_{f}}{M_{i}}[p_{M}^{7}]\\
\Gamma(3^{-} \rightarrow 1^{-})= C_{M}\frac{16g_{YH}^{2}}{105\pi
f_{\pi}^{2}\Lambda^{4}} \frac{M_{f}}{M_{i}}[p_{M}^{7}]
\end{gather}

Here $M_{i}$, $M_{f}$ gives initial and final momentum, $\Lambda$ is chiral symmetry breaking scale of 1 GeV. $p_{M}$, $m_{M}$ denotes the final momentum and mass of light pseudoscalar meson. The coupling constant plays a key role in the phenomenology study of heavy-light mesons. These dimensionless coupling constants describe the strength of transition between H-H field (negative-negative parity), S-H field (positive-negative parity), T-H field (positive-negative parity). These couling constants are notated as $g_{HH}$, $g_{SH}$, $g_{TH}$, $g_{XH}$, $g_{YH}$ etc. The coefficient $C_{M}$ for different pseudoscalar particles are: $C_{\pi^{\pm}}$, $C_{K^{\pm}}$, $C_{K^{0}}$, $C_{\overline{K}^{0}}=1$, $C_{\pi^{0}}=\frac{1}{2}$ and $C_{\eta}=\frac{2}{3}(c\bar{u}, c\bar{d})$ or $\frac{1}{6}(c\bar{s})$. In our paper, we are not including higher order corrections of $\frac{1}{m_{Q}}$ to introduce new couplings. We also expect that higher corrections give a small contribution in comparison the leading order contributions. The numerical values of various meson masses used in the calculations are listed in Table \ref{tab:input}.
\setlength{\tabcolsep}{0.09em} %
{\renewcommand{\arraystretch}{0.2}%

\begin{table*}{\normalsize
\renewcommand{\arraystretch}{1.0}
\tabcolsep 0.2cm \caption{\label{tab:input}Numerical value of the meson masses used in this work \cite{8}.}
 \noindent
\begin{tabular}{cccccc}
  \hline
  \hline
  States &$B^{0}$&$B^{\pm}$&$B^{*}$&$B_{s}$&$B^{*}_{s}$\\
  \hline
  Masses(MeV)&5279.58&5279.25&5325.20&5366.77&5415.40\\
  \hline
  States &$\pi^{\pm}$&$\pi^{0}$&$\eta$&$K^{+}$&$K^{0}$\\
  \hline
  Masses(MeV)&139.57&134.97&547.85&493.67&497.61\\
\hline \hline
\end{tabular}}
\end{table*}
\section{Numerical analysis}

Assigning a particular $J^P$ (spin-parity quantum number) to the experimentally predicted excited states is very crucial. A specific position of the state in its mass spectra can help in redeeming many other important strong interaction parameters such as hadronic coupling constants, branching ratios, spins and mass splittings, decay widths and many more. Therefore, in this analysis, we aim to suggest a particular $J^P$ state to the recently strange bottom states observed by LHCb. Moreover, we also completed the empty spaces of this strange mass spectra by predicting the masses and other parameters of the missing bottom spectra in the framework of HQET. \\ In the bottom strange sector, only ground states 1S ($0^-, 1^-$) states are confirmed both experimentally and theoretically. The rest of the spectra is still unknown. Although many other theoretical models, such as the chiral perturbation theory, the chiral unitary approach, the Regge trajectory, the $^3P_0$ model, the relativistic flux tube model, QCD sum rules, and the lattice QCD, etc tried to fill the gaps but yet they are not verified experimentally \cite{13,14,15,16,17,18,19,20,21,22,23,24,25,26,27,28,29,30,31}. In addition to this, such theoretically calculated masses are unreliable because all these theoretical models depend on certain unknown parameters. The HQET framework chosen by us is free from such parameters and uses the most important relations of spin and flavour symmetry present in heavy-light mesons.
\subsection{Mass spectroscopy:} To study the behavior of heavy-light mesons, masses are the prime property that determines many other properties of the mesons. Therefore, we start our calculations by predicting the masses of unavailable strange bottom states using the flavour and spin symmetry property $ \Delta_{F}^{(c)} =\Delta_{F}^{(b)}$ and $ \lambda_{F}^{(c)} = \lambda_{F}^{(b)}$. As discussed in Section II, the parameter $\Delta_{F}$ is defined in terms of the spin averaged mass splittings between the higher state doublet and ground state doublet, whereas the other parameter $\lambda_{F}$ is the mass splittings between the spin partners of the doublets. \\
The main aim of our calculation is to incorporate the $J^P$ to recently observed strange states of the bottom flavour $B_{sJ}(6063)$ and $B_{sJ}(6114)$ and to fill the gaps near these states so that our calculations can provide motivation and reference to the yet to be predicted experimental information. The details of the numerical analysis on the mass spectrum are as follows: \\
\begin{table}[ht]
        \centering
       \caption{\label{table2} Input values used in this work. All values are in units of $MeV$.}
          \begin{tabular}{|c|c|c|c|}
      \hline
     State & $J^{P}$ & $ c\overline{s}$ &$ b\overline{s}$ \\
     \hline
     $1^{1}S_{0}$ & $0^{-}$ &  1968.35 \cite{6}  &  5366.92 \cite{6}  \\
     \hline
      $1^{3}S_{1}$ & $1^{-}$ & 2112.20 \cite{6} & 5415.40 \cite{6} \\
      \hline
      $1^{3}P_{0}$ & $0^{+}$ & 2317.80 &  -  \\
      \hline
      $1^{1}P_{1}$ & $1^{+}$ & 2459.50 &  -  \\
      \hline
      $1^{3}P_{1}$ & $1^{+}$ & 2535.11 & -  \\
      \hline
      $1^{3}P_{2}$ & $2^{+}$ & 2569.10 & -  \\
      \hline
      $1^{3}D_{1}$ & $1^{-}$ &  2859.00 & -   \\
     \hline
      $1^{1}D_{2}$ & $2^{-}$ & 2902.40 & - \\
      \hline
      $1^{3}D_{2}$ & $2^{-}$ &  2895.40 & -  \\
     \hline
      $1^{3}D_{3}$ & $3^{-}$ & 2860.50 & - \\
      \hline
      $2^{1}S_{0}$ & $0^{-}$ &  2676.20 & -  \\
     \hline
      $2^{3}S_{1}$ & $1^{-}$ & 2714.00 & - \\
      \hline
       \end{tabular}
    \label{tab:my_label1}
\end{table}
Using the masses of charm and bottom states tabulated in Table \ref{table2}, the spin averaged mass splittings $\Delta_{F}^{(c)}$ and the hyperfine splittings $\lambda_{F}^{(c)}$ for $1S, 1P, 1D$ and $2S$ comes out to be\\
\begin{center}
$\Delta_{\tilde{H}}^{(c)}$ = 628.313 $MeV$\\
$\Delta_{S}^{(c)}$ = 347.838 $MeV$\\
$\Delta_{T}^{(c)}$ = 480.116 $MeV$\\
$\Delta_{X}^{(c)}$ = 763.513 $MeV$\\
$\Delta_{Y}^{(c)}$ = 798.804 $MeV$\\
$\lambda_{\tilde{H}}^{(c)}$ = $(159.589)^{2} MeV^{2}$\\
$\lambda_{S}^{(c)}$ = $(290.892)^{2} MeV^{2}$\\
$\lambda_{T}^{(c)}$ = $(180.360)^{2} MeV^{2}$\\
$\lambda_{X}^{(c)}$ = $(181.228)^{2} MeV^{2}$\\
$\lambda_{Y}^{(c)}$ = $(204.573)^{2} MeV^{2}$
\end{center}
\setlength{\tabcolsep}{0.09em} %
{\renewcommand{\arraystretch}{0.2}%
\begin{table*}[h]{\normalsize
\renewcommand{\arraystretch}{1.0}
\tabcolsep 0.2cm \caption{\label{strangebottom}Predicted values of
the radially excited strange $2S, 1P$ and $1D$ bottom meson states. All the
masses are in $MeV$ units}
 \noindent
 \begin{tabular}{ccccccc}
\hline
 & Without
 corrections &Corrections
 in $\lambda_{F}$&Corrections
 in
 $\Delta_{F}$ &  Correction in both parameters $\lambda_{F}$ and
 $\Delta_{F}$\\
 \hline\hline
$ 0^{-}(2^{1}S_{0})$&6018.92&6020.98&6017.74&6019.80\\
 $1^{-}(2^{3}S_{1}) $&6035.82&6035.13&6034.65&6033.96\\
$ 0^{+}(1^{3}P_{0})$&5706.86&5714.07&5705.35&5712.56\\$ 1^{+}(1^{1}P_{1})$
&5765.87&5763.47&5764.37&5761.97\\
$ 1^{+}(1^{3}P_{1})$ &5874.18&5875.68&5867.30&5868.80\\
$2^{+}(1^{3}P_{2})$&5888.93&5888.03&5882.06&5881.16\\
$ 1^{-}(1^{3}D_{1})$&6175.67&6174.23&6180.17&6178.73\\
$ 2^{-}(1^{1}D_{2})$&6161.47&6162.33&6165.98&6166.84\\
$ 2^{-}(1^{3}D_{2})$ &6211.53&6205.10&6210.70&6209.16\\
$ 3^{-}(1^{3}D_{3})$ &6195.34&6194.50&6194.51&6195.60\\
\hline
\end{tabular}}
\end{table*}
The charm mesons with $J^P = 2^-$ of D-wave with j = 3/2 and j = 5/2 for $1^1D_2$ and $1^3D_2$ are experimentally unavailable, so we have taken the average of the theoretical masses \cite{13,47,48,49} for them.\\ Masses obtained for $B_s$ mesons with the help of the symmetries are listed in the $2^{nd}$ column of Table \ref{strangebottom}. Masses predicted for P-wave j = 3/2 strange bottom states $1^{3}P_{1}$ and $1^{3}P_{2}$ are in very good agreement with the experimental masses predicted for these states by LHCb \cite{37}, CDF \cite{11,35} and D0 \cite{36}. Our calculated masses for $1^{3}P_{1}$ and $1^{3}P_{2}$ are only 0.78\% , 0.84\%  deviating from their experimental values respectively.
In comparison with predictions of other theoretical models shown in Table \ref{strangebottom1}, our calculated masses are in good agreement. Note that predictions of Ref.\cite{13,51} are smaller by 100 $MeV$ than our results. While data in \cite{40} is larger than our masses by order of 100 $MeV$. Also, the calculated mass for $J^P = (1^-, 2^-), 3^-$ belonging to D-wave with j = 3/2, j = 5/2 and $J^P = 1^-$ of radially excited S-wave have also beautifully matched with the LHCb states $B_{sJ}(6114)$ and $B_{sJ}(6063)$ respectively. 
Since masses of these two states $B_{sJ}(6114)$ and $B_{sJ}(6063)$ deviates only by 61, 47, 81 $MeV$ and 28 $MeV$ from our predicted values, so one easily conclude that $B_{sJ}(6063)$ state belong to $J^P$ $2S1^-$ while the other state $B_{sJ}(6114)$ belong to one of the $J^P$ of 1 D-wave. Authors in Ref.\cite{40} predicted the $J^P$ for $B_{sJ}(6114)$ as $1D 1^-$ for j = 3/2.  But, to prove convincingly, we computed the masses by taking higher order corrections as splitting parameters ($\Delta_{F}$ and $ \lambda_{F}$) can drastically change in the presence of QCD and higher order ($\tilde{}{1/m_Q}$) corrections in the HQET lagrangian. Also, the calculated masses should be able to predict other parameters, such as decay width, which should match with experimental data. Therefore, in the next part, we present the analysis of such corrections to these splittings and calculate the two-body strong decay widths of these bottom states. 
\\  QCD and higher order ($\tilde{1/m_Q}$) corrections are applied to a scale of $\Lambda_{QCD}/m_Q$, where they can significantly decide the level of symmetry breaking. The corrections to $\Delta_{F}$ and $\lambda_F$ parameters change heavy quark symmetry relations to $ \Delta_{F}^{(b)} =\Delta_{F}^{(c)} + \delta\Delta_F$ and $ \lambda_{F}^{(b)} = \lambda_{F}^{(c)}\delta\lambda_F$. We apply such corrections to these splittings one by one and check the effect of these corrections on the bottom masses followed by a step in which both corrections would be applied simultaneously. In the case of $\lambda_F$ parameter, QCD corrections are dominant over $1/m_Q$ corrections because these mass splitting parameters $\lambda_F$ originate from the chromomagnetic interactions. The leading QCD corrections to the $\lambda_F$ are in the form of $\lambda_{F}^{(b)} = \lambda_{F}^{(c)}{{(\frac{\alpha_s(m_b)}{\alpha_s(m_c)})}}^{9/25}$.
The parameters $\alpha_s(m_b)$ and $\alpha_s(m_c)$ for applying the QCD corrections to this splitting parameters are taken as 0.22 and 0.36 \cite{50}. The corrections in $\lambda_F$ parameters modify the value to:
\begin{center}
$\lambda_{\tilde{H}}^{(c)}$ = $(133.662)^{2} MeV^{2}$\\
$\lambda_{S}^{(c)}$ = $(243.632)^{2} MeV^{2}$\\
$\lambda_{T}^{(c)}$ = $(151.058)^{2} MeV^{2}$\\
$\lambda_{X}^{(c)}$ = $(151.785)^{2} MeV^{2}$\\
$\lambda_{Y}^{(c)}$ = $(171.33)^{2} MeV^{2}$\\
\end{center}
The $B_s$ masses inherited using these corrections are tabulated in the $3^{rd}$ column of Table \ref{strangebottom}. The resulting masses are deflected upto 7.5 $MeV$ from their initial masses and have resulted in reducing the gap between our and LHCb masses.  
Now we analyze the effect of $1/m_Q$ and QCD corrections to our other parameter $\Delta_F$, which depicts the mass splittings between the higher mass doublet and the ground state H field states. The corrections to these parameters are in the form of $\delta\Delta_F$, where $F = S, T, X, Y, \tilde{H}$. The best-fitted values of these corrections come to be $\mathcal{O}$ (0.63) $GeV$, $\mathcal{O}$(0.35) $GeV$, $\mathcal{O}$(0.49) $GeV$, $\mathcal{O}$(0.77) $GeV$ and $\mathcal{O}$(0.80) $GeV$ for doublets $2S(0^{-},1^{-})$, $1P(0^{+},1^{+})$, $1P(1^{+},2^{+})$, $1D(1^{-},2^{-})$ and $1D(2^{-},3^{-})$ respectively. Masses are calculated using these corrections are tabulated in $4^{th}$ column of Table \ref{strangebottom}. A comparison of masses concluded that the deviation lies in the range of 0.83 - 6.90 $MeV$, which again shows that masses are not affected much by corrections too but again resulted in narrowing the gap between our and experimental masses. Lastly, applying both corrections simultaneously the masses obtained are listed in the last column of Table \ref{strangebottom}. It can be summarized that the effect of such corrections is very small in the higher excited states. Because of the higher angular momentum, such states do not remain in their stable state for much time and thus do not explicitly show chromomagnetic effects. Although such corrections have resulted in narrowing the gap between the predicted and experimentally predicted mass values but have not adversely affected the masses. To suggest a particular $J^P$ value for LHCb observed strange bottom states $B_{sJ}(6063)$ and $B_{sJ}(6114)$, we explore the second utmost important property of heavy-light mesons i.e. decay widths. 

\begin{table*}[h]{\normalsize
\renewcommand{\arraystretch}{1.0}
\tabcolsep 0.2cm \caption{\label{strangebottom1}The predicted values of bottom-strange meson masses ($MeV$) compared with some other model predictions.}
 \noindent
 \begin{tabular}{ccccccc}
\hline
 & Ours
  &Ref.\cite{51}
 & Ref.\cite{13} & Ref.\cite{40}\\
 \hline\hline
$ 0^{-}(2^{1}S_{0})$&6018.92&6003&5985&6025\\
 $1^{-}(2^{3}S_{1}) $&6035.82&6029&6019&6033\\
$ 0^{+}(1^{3}P_{0})$&5706.86&5812&5804&5709\\
$ 1^{+}(1^{1}P_{1})$
&5765.87&5828&5805&5768\\
$ 1^{+}(1^{3}P_{1})$ &5874.18&5842&5842&5875\\
$2^{+}(1^{3}P_{2})$&5888.93&5840&5820&5890\\
$ 1^{-}(1^{3}D_{1})$&6175.67&6119&6127&6247\\
$ 2^{-}(1^{1}D_{2})$&6161.47&6128&6095&6256\\
$ 2^{-}(1^{3}D_{2})$ &6211.53&6157&6140&6292\\
$ 3^{-}(1^{3}D_{3})$ &6195.34&6172&6103&6297\\
\hline
\end{tabular}}
\end{table*}

\subsection{Strong Decays}
We apply the effective Lagrangian
approach discussed in Sec II to calculate the OZI allowed
two body strong decay widths and the various branching
ratios involved with the bottom states $B_s(2S)$, $B_s(1P)$ and $B_s(1D)$. The numerical value of the partial and total decay widths of these states are given in Table \ref{width1} and \ref{width2}. Here, we need to emphasize that calculated total decay widths for the above states do not include the contribution of decays with the emission of vector mesons ($\omega,\rho,K^*,\phi$). Since the contribution of vector mesons to total decay widths is small compared to pseudoscalar mesons. They give the contribution of $\pm 10 MeV$ \cite{15} to total decay widths for above analyzed states.
\\To choose the possible $J^P$ for the LHCb bottom state $B_{sJ}(6114)$, we calculated the total decay width for all the possible $J^P$'s 
$1^3D_1$, $1 ^1D_2$ and $1 ^3D_3$ in terms of strong coupling constants. Values come out to be \begin{center}
$\Gamma(1^3D_1)$ = 6052.34${g}^{2}_{XH}$\\
$\Gamma(1 ^1D_2)$ = 4416.36${g}^{2}_{XH}$\\
$\Gamma(1 ^3D_3)$ = 933.75${g}^{2}_{YH}$\\
\end{center}
On comparing these calculated decay widths with the experimental value of decay width of 66 $MeV$, the coupling constant values come out to be
\begin{center}
     ${g}_{XH} = 0.104$\\ 
    ${g}_{XH} = 0.12$\\
    ${g}_{YH} = 0.26$
\end{center}

The available theoretical values of $g_{YH}$ are 0.61 \cite{54}, 0.53 \cite{25}, 0.42 \cite{53}. Our computed value of $g_{YH}$ is much smaller, ruling out $1^3D_3$ state from possible $J^P$'s for state $B_{sJ}(6114)$. Now, we are left with two available $J^P$ ($1^3D_1$ and $1 ^1D_2$). In literature, the available theoretical values of $g_{XH}$ are 0.41 \cite{28}, 0.45 \cite{32}, 0.53 \cite{52}, 0.19 \cite{25}. In Ref.\cite{25}, the coupling constant $g_{XH}$ is calculated for state $B(5970)^0$ is $0.19\pm 0.049$. They assigned $1^3D_1$ for this state and computed coupling constant by comparing their theoretical decay width with the experimental value. We are also following the same procedure, and obtained value of $g_{XH}$ enables us to assign the $J^P$ state of $B_{sJ}(6114)$. The coupling value for $1^3D_1$ and $1^1D_2$ obtained are consistent with $g_{XH} = 0.19$. We suggest these two $J^P$ $1^3D_1$ and $1^1D_2$ as most favorable for state $B_{sJ}(6114)$. But $B_{sJ}(6114)$ were observed in BK modes, so it cannot have $J^P = 2^-$. Since it does not satisfy the conservation of parity and angular momentum simultaneously for $J^P = 2^-$. So, we are assigning a particular $J^P = 2^- (1^3D_1)$ to state $B_{sJ}(6114)$. Further experimental investigations like branching ratios may provide the necessary conclusions.
\subsubsection{$1 D$ state}
The natural parity D states  $1 ^3D_1$ and $1 ^3D_3$ both are dominant in BK decay mode with branching fractions of 56.73\% and 36.67\% respectively, while the unnatural parity states $1 ^1D_2$ and $1 ^3D_2$ shows dominance in $B^*K$ decay channel with branching fractions of 77.51\% and 67.02\% respectively. Column 4 of Table \ref{width1} gives the ratio of the partial decay widths for $1D$ bottom states with respect to its
partial decay width $B^{*+}K^{-}$. Apart from the decay channels
listed in this Table \ref{width1}, these bottom states also decay to P-wave bottom meson states, which occur via D-wave and thus due to the small phase space, these decay modes are suppressed when compared to decays to ground state S-wave mesons and hence are not shown in Table \ref{width1}. The calculated value of $g_{XH} = 0.12$ can be beneficial in finding the total and partial decay width of unobserved bottom state $1 ^3D_1$. Thus the calculated total decay width for this state is 87.15 $MeV$ which deviates by 22.87\% from  the result of Ref.\cite{39} 
\subsubsection{$1 P$ state}
We have also analyzed the strong decay widths $1P$ bottom states and calculated the various branching ratios involved. The calculated value of the partial decay widths for the bottom states $1 ^3P_0$, $1 ^1P_1$, $1 ^3P_1$ and $1 ^3P_2$ are
listed in Table \ref{width2}. The state $1^3P_0 $ decaying to BK decay channel is kinematically suppressed as the predicted mass for this state is lower than the BK threshold value. This is a similar situation as seen in the charm sector for the same $J^P$ state $D^*_{s0}(2317)$, thus reflecting the flavour symmetry in heavy hadrons. The only kinematically allowed decay channel for this state is $B_s\pi$. Its spin partner $1 ^1P_1$ is also decaying to $B^*_s\pi^0$ decay mode only while all other modes are suppressed. Using the available value of coupling constant $g_{SH} = 0.56$ \cite{25,32,40}, predicted decay widths for $(0^+,1^+)$ doublet are calculated as 
\begin{center}
    $\Gamma(1^3P_0)$ = 46.25 $MeV$\\
    $\Gamma(1^1P_1)$ = 50.51 $MeV$\\
\end{center}
Our estimated values are highly overestimated with the results of Ref.\cite{40} and underestimated with the data of QPC model, and chiral quark model \cite{28,36,40}. \\The study of the doublet $(1^+,2^+)$ shows that $B^*_s\pi^0$ and $B_s\pi^0$ are the dominant decay modes for $1 ^3P_1$ and $1 ^3P_2$ states with branching ratio of 85.33\% and 32.30\% respectively. 
Total decay width calculated by taking a sum of its partial decay widths listed in Table \ref{width2}
\begin{center}
    $\Gamma(1^3P_1)$ = 14.33 $MeV$\\
    $\Gamma(1^3P_2)$ = 26.74 $MeV$\\
    These decay values depict these states to be narrower but are still overestimated with their experimental values measured by LHCb and CDF collaborations \cite{11,37} but are in good agreement with the theoretical data \cite{11,40} where authors used mixing angle to calculate the width of $1 ^3P_1$ state.
\end{center}
\subsubsection{$2 S$ state}
For the radially ground state S-wave bottom states, Table \ref{width2}  reveals $B^{*}K^{+}$ mode to be the dominant decay
 mode both for $\widetilde{B}_{1s}^{*}$ and $\widetilde{B}_{0s}$ bottom states with branching fraction of 22.53$\%$
 and 33.46$\%$, respectively.  Hence the decay modes
 $B^{*}K^{+}$ is suitable for the experimental search for the missing
 strange $2S$ bottom meson states. Using the strong coupling value $\widetilde{g_{HH}} = 0.31$ \cite{53}, total decay widths for bottom state $\widetilde{B}_{1s}^{*}$ and its spin partner $\widetilde{B}_{0s}$ are predicted as 207.44 $MeV$ and 231.11 $MeV$ respectively. These speculated values indicate these radial bottom states to have broad resonance, which is in good agreement with the results of Ref.\cite{11,40}.
Apart from the mentioned partial decay widths, these bottom states also decay to D-wave bottom mesons. But these decays are suppressed in our calculations because of their small contributions.

 \begin{table}
       \caption
{Strong decay width of 
strange bottom mesons $B_s(2S0^-)$, $B_s(2S1^-)$, $B_s(1P0^+)$, $B_s(1P1^+)$, $B_s(1P1^+)$, $B_s(1P2^+)$. Ratio in 5th column
represents the $\widehat{{\bf \Gamma}}=
\frac{\Gamma}{\Gamma(B_{sJ}^{*} \rightarrow B^{*0}K^{+})}$. Fraction gives the percentage of the partial decay
width with respect to the total decay width. }\label{width2}
\begin{center}
\begin{tabular}{ c | c | c | c | c }
\hline \hline $nLs_{l}J^{P}$&Decay channel&Decay
Width($MeV$)&Ratio&Fraction\\
\hline
2$S_{s1/2}0^{-}$&$B^{*0}K^{0}$&785.160$\widetilde{g}^{2}_{HH}$&1&32.64\\
&$B^{*+}K^{-}$&804.921$\widetilde{g}^{2}_{HH}$&1.02&33.46\\
&$B^{*}_{s}\pi^{0}$&737.438$\widetilde{g}^{2}_{HH}$&0.93&30.66\\
&$B^{*}_{s}\eta$&77.419$\widetilde{g}^{2}_{HH}$&0.09&3.21\\
&Total&2404.940$\widetilde{g}^{2}_{HH}$&&\\
\hline
$2S_{s1/2}1^{-}$&$B^{0}K^{0}$&342.716$\widetilde{g}^{2}_{HH}$&0.72&15.87\\
&$B^{+}K^{-}$&350.669$\widetilde{g}^{2}_{HH}$&0.73&16.24\\
&$B_{s}\pi^{0}$&292.676$\widetilde{g}^{2}_{HH}$&0.61&13.55\\
&$B_{s}\eta$&58.29$\widetilde{g}^{2}_{HH}$&0.12&2.70\\
&$B^{*0}K^{0}$&474.96$\widetilde{g}^{2}_{HH}$&1&22.00\\
&$B^{*+}K^{-}$&486.477$\widetilde{g}^{2}_{HH}$&1.2&22.53\\
&$B^{*}_{s}\pi^{0}$&466.530$\widetilde{g}^{2}_{HH}$&0.98&21.61\\
&$B^{*}_{s}\eta$&36.988$\widetilde{g}^{2}_{HH}$&0.07&1.71\\
&Total&2158.650$\widetilde{g}^{2}_{HH}$&&\\
\hline
$1P_{s1/2}0^{+}$&$B_{s}\pi^{0}$&147.500${g}^{2}_{SH}$&-&100\\
&$B_{s}\eta$&-&&-\\
&$B^{+}K^{0}$&-&-&-\\
&$B^{-}K^{+}$&-&-&-\\
&Total&147.500${g}^{2}_{SH}$&&\\
\hline
1$P_{s1/2}1^{+}$&$B^{*}_{s}\pi^{0}$&161.080${g}^{2}_{SH}$&-&-\\
&$B_{s}^{*}\eta$&-&-&-
\\&$B^{*+}K^{0}$&-&-&-\\
&$B^{*-}K^{+}$&-&-&-\\
&Total&1161.080${g}^{2}_{SH}$&&\\
 \hline
1$P_{s13/2}1^{+}$&$B^{*0}K^{0}$&5.970${g}^{2}_{TH}$&1&6.66\\
&$B^{*+}K^{-}$&7.12${g}^{2}_{TH}$&1.19&7.93\\
&$B^{*}_{s}\pi^{0}$&76.49${g}^{2}_{TH}$&12.81&85.37\\
&$B^{*}_{s}\eta$&-&-&-\\
&Total&89.59${g}^{2}_{TH}$&&\\
\hline
1$P_{s13/2}2^{+}$&$B^{0}K^{0}$&18.48${g}^{2}_{TH}$&22.67&11.05\\
&$B^{+}K^{-}$&20.13${g}^{2}_{TH}$&2.91&12.04\\
&$B_{s}\pi^{0}$&59.590${g}^{2}_{TH}$&8.62&35.65\\
&$B_{s}\eta$&-&-&-\\
&$B^{*0}K^{0}$&6.92${g}^{2}_{TH}$&1&4.13\\
&$B^{*+}K^{-}$&7.92${g}^{2}_{TH}$&1.14&4.73\\
&$B^{*}_{s}\pi^{0}$&54.09${g}^{2}_{TH}$&7.82&32.36\\
&$B^{*}_{s}\eta$&-&-&-\\
&Total&167.15${g}^{2}_{TH}$&&\\
\hline

 \hline
\hline
\end{tabular}
\end{center}
\end{table}


\begin{table}
       \caption
{Strong decay width of 
strange bottom mesons  $B_s(1D1^-)$, $B_s(1D2^-)$, $B_s(1D2^-)$ and $B_s(1D3^-)$. Ratio in 5th column
represents the $\widehat{{\bf \Gamma}}=
\frac{\Gamma}{\Gamma(B_{sJ}^{*} \rightarrow B^{*0}K^{+})}$. Fraction gives the percentage of the partial decay
width with respect to the total decay width. }\label{width1}
\begin{center}
\begin{tabular}{ c | c | c | c | c }
\hline \hline $nLs_{l}J^{P}$&Decay channel&Decay
Width($MeV$)&Ratio&Fraction\\
\hline
1$D_{s3/2}1^{-}$&$B^{0}K^{0}$&1706.81${g}^{2}_{XH}$& 2.71&28.20\\
&$B^{+}K^{-}$&1727.24${g}^{2}_{XH}$&2.74&28.53\\
&$B_{s}\pi^{0}$&863.737${g}^{2}_{XH}$&1.37&14.27\\
&$B_{s}\eta$&126.531${g}^{2}_{XH}$&0.20&2.09\\
&$B^{*0}K^{0}$&628.449${g}^{2}_{XH}$&0.51&5.35\\
&$B^{*+}K^{-}$&635.727${g}^{2}_{XH}$&0.06&0.66\\
&$B^{*}_{s}\pi^{0}$&323.847${g}^{2}_{XH}$&1&10.38\\
&$B^{*}_{s}\eta$&40.0077${g}^{2}_{XH}$&1.01&10.50\\
&Total&6052.34${g}^{2}_{XH}$&&\\
\hline
1$D_{s3/2}2^{-}$&$B^{*0}K^{0}$&1701.43${g}^{2}_{XH}$&1&38.52\\
&$B^{*+}K^{-}$&1722.22${g}^{2}_{XH}$&1.01&38.99\\
&$B^{*}_{s}\pi^{0}$&889.31${g}^{2}_{XH}$&0.52&20.13\\
&$B^{*}_{s}\eta$&103.31${g}^{2}_{XH}$&0.06&2.34\\
&Total&4416.36${g}^{2}_{XH}$&&\\
 \hline
1$D_{s5/2}2^{-}$&
$B^{*0}K^{0}$&290.97${g}^{2}_{YH}$&1&33.08\\
&$B^{*+}K^{-}$&298.45${g}^{2}_{YH}$&1.02&33.93\\
&$B^{*}_{s}\pi^{0}$&251.13${g}^{2}_{YH}$&0.868&28.55\\
&$B^{*}_{s}\eta$&38.85${g}^{2}_{YH}$&0.13&4.41\\
&Total&879.41${g}^{2}_{XH}$&&\\
\hline
 1$D_{s5/2}3^{-}$&$B^{0}K^{0}$&168.95${g}^{2}_{YH}$&1.21&18.09\\
&$B^{+}K^{-}$& 173.47${g}^{2}_{YH}$&1.24&18.57\\
&$B_{s}\pi^{0}$&140.16${g}^{2}_{YH}$&18&15.01\\
&$B_{s}\eta$&27.27${g}^{2}_{YH}$&0.19&2.92\\
&$B^{*0}K^{0}$&139.01${g}^{2}_{YH}$&1&14.88\\
&$B^{*+}K^{-}$&142.78${g}^{2}_{YH}$&1.02&15.29\\
&$B^{*}_{s}\pi^{0}$&125.11${g}^{2}_{YH}$&0.90&13.39\\
&$B^{*}_{s}\eta$&16.98${g}^{2}_{YH}$&0.12&1.81\\
&Total&933.75${g}^{2}_{YH}$&&\\

 \hline
\hline
\end{tabular}
\end{center}
\end{table}

 
\section{Conclusion}
In this paper, we have applied the Heavy Quark
Effective Theory to examine the recently
observed strange bottom mesons, $B_{sJ}(6063)$ and $B_{sJ}(6114)$ by LHCb collaborations. In this framework, we have calculated masses of strange bottom meson $2S, 1P, 1D$ with the use of available experimental and theoretical data on charm mesons and including non-perturbative parameters ($\Delta_{F}$ and $ \lambda_{F}$). These predicted masses for the above-said states have beautifully matched with other models predictions. Also, by taking ${1/m_Q}$ corrections in terms of $\delta\Delta_F$ and $\delta\lambda_F$, we estimated masses of strange bottom mesons $2S, 1P, 1D$, which narrow the gap between our results and experimental data. On the basis of computed masses, we have identified strange bottom states $B_{sJ}(6063)$ as $2^{3}S_{1}$ and give three possible $J^P$'s ( $1^-, 2^-, 3^-$) belonging D-wave to $B_{sJ}(6114)$ state. We have analyzed strong decay widths for possible $J^P$'s state for $B_{sJ}(6114)$ and concluded that most favourable states for $B_{sJ}(6114)$ are $1^{3}D_{1}$. In addition to this, we have predicted the branching ratios and the coupling constants for the above states, which can provide crucial information for future experimental searches.
\section{Acknowledgment}
The authors thankfully acknowledge the financial support by the
Department of Science and Technology (SERB/F/9119/2020), New Delhi.


\begin{thebibliography}{}
\bibitem{1} P. del Amo Sanchez et al. (BABAR Collaboration), Phys. Rev. D 82, 111101 (2010).
\bibitem{2}R. Aaij et al. (LHCb Collaboration), J. High Energy Phys. 09 (2013) 145.
\bibitem{3} R. Aaij et al. (LHCb Collaboration), Phys. Rev. D 91, 092002 (2015).
\bibitem{4} B. Aubert et al. (BABAR Collaboration), Phys. Rev. Lett. 97, 222001 (2006).
\bibitem{5} J. Brodzicka et al. (Belle Collaboration), Phys. Rev. Lett. 100, 092001 (2008).
\bibitem{6} R. Aaij et al. (LHCb Collaboration), J. High Energy Phys. 02 (2016) 133.
\bibitem{7} R. Aaij et al. (LHCb Collaboration), Phys. Rev. Lett. 126, 122002 (2021).
\bibitem{8}  R.L. Workman et al. [Particle Data Group], Prog. Theor. Exp. Phys. 2022, 083C01 (2022)
\bibitem{9} R. Aaij et al. (LHCb Collaboration), Eur. Phys. J. C 81, 601 (2021).
\bibitem{10} M. Acciarri et al. (L3 Collaboration), Phys. Lett. B 465, 323 (1999).
\bibitem{11} T. A. Aaltonen et al. (CDF Collaboration), Phys. Rev. D 90, 012013 (2014).
\bibitem{12} R. Aaij et al. (LHCb Collaboration), J. High Energy Phys. 04 (2015) 024.
\bibitem{13} M. Di Pierro and E. Eichten,  Phys. Rev. D 64, 114004 (2001).
\bibitem{14} Y. Sun, Q. T. Song, D. Y. Chen, X. Liu, and S. L. Zhu, Phys. Rev. D 89, 054026 (2014).
\bibitem{15} S. Godfrey, K. Moats, and E. S. Swanson, Phys. Rev. D 94, 054025 (2016).
\bibitem{16} Q. F. Lu, T. T. Pan, Y. Y. Wang, E. Wang, and D. M. Li, Phys. Rev. D 94, 074012 (2016).
\bibitem{17} I. Asghar, B. Masud, E. S. Swanson, F. Akram, and M. A. Sultan, Eur. Phys. J. A 54, 127 (2018).
\bibitem{18} S. Godfrey and K. Moats, Eur. Phys. J. A 55, 84 (2019).
\bibitem{19} Z. H. Wang, Y. Zhang, T. H. Wang, Y. Jiang, Q. Li, and G. L. Wang, Chin. Phys. C 42, 123101 (2018).
\bibitem{20} G. L. Yu and Z. G. Wang, Chin. Phys. C 44, 033103 (2020).
\bibitem{21} H. A. Alhendi, T.M. Aliev, and M. Savcı, J. High Energy Phys. 04 (2016) 050.
\bibitem{22} J. Ferretti and E. Santopinto, Phys. Rev. D 97, 114020 (2018).
\bibitem{23} Z. G. Wang, Eur. Phys. J. C 74, 3123 (2014).
\bibitem{24} H. Xu, X. Liu, and T. Matsuki, Phys. Rev. D 89, 097502 (2014).
\bibitem{25} Z. G. Wang, Eur. Phys. J. Plus 129, 186 (2014).
\bibitem{26} J. M. Zhang and G. L. Wang, Phys. Lett. B 684, 221 (2010).
\bibitem{27} Z. G. Luo, X. L. Chen, and X. Liu, Phys. Rev. D 79, 074020 (2009).
\bibitem{28} P. Gupta and A. Upadhyay, Phys. Rev. D 99, 094043 (2019).
\bibitem{29} S. L. Zhu and Y. B. Dai, Mod. Phys. Lett. A 14, 2367 (1999).
\bibitem{30} A. H. Orsland and H. Hogaasen, Eur. Phys. J. C 9, 503 (1999).
\bibitem{31} G. L. Yu, Z. G. Wang, and Z. Y. Li, Eur. Phys. J. C 79, 798 (2019).
\bibitem{32} P. Colangelo, F. De Fazio, F. Giannuzzi, and S. Nicotri, Phys. Rev. D 86, 054024 (2012).
\bibitem{33} X.-H. Zhong and Q. Zhao, Phys. Rev. D 78, 014029 (2008).
\bibitem{34} Stephen Godfrey and Kenneth Moats, Eur. Phys. J. A 55, 84 (2019).

\bibitem{35} T. Aaltonen et al. (CDF Collaboration), Phys. Rev. Lett. 100, 082001 (2008).
\bibitem{36}V. M. Abazov et al. (D0 Collaboration), Phys. Rev. Lett. 100, 082002 (2008).
\bibitem{37} R. Aaij et al. (LHCb Collaboration), Phys. Rev. Lett. 110, 151803 (2013).
\bibitem{38} Bing Chen, Si-Qiang Luo, Ke-Wei Wei,and Xiang Liu, Phy. Rev. D 105, 074014 (2022).
\bibitem{39} Qi Li, Ru-Hui Ni, and Xian-Hui Zhong, Phy. Rev. D 103, 116010 (2021).
\bibitem{40} Keval Gandhi et al.,Eur. Phys. J. C 82, 777 (2022).
\bibitem{41} M. Neubert, Phys. Rept.245,259 (1994).
\bibitem{42} A. F. Falk and T. Mehen, Phys. Rev. D 53 231 (1996).

\bibitem{43} M. B.Wise, Phys. Rev. D 45, R2188 (1992).
\bibitem{44} G. Burdman and J. F. Donoghue, Phys. Lett. B 280, 287
(1992).
\bibitem{45} P. L. Cho, Phys.Lett. B 285, 145 (1992).
\bibitem{46} R. Casalbuoni, A. Deandrea, N.Di Bartolomeo, R.Gatto,
F.Feruglio and G.Nardulli, Phys. Lett. B 299, 139 (1993).
\bibitem{47} V.Kher et al., Chinese Phys. C 41, 073101 (2017).
\bibitem{48} D. Ebert, R. N. Faustov, and V. O. Galkin, Eur. Phys. J. C 66, 197 (2010).
\bibitem{49} S. Godfrey and K. Moats, Phys. Rev. D 93, 034035 (2016).
\bibitem{50} G. Amoros, M. Beneke and M. Neubert, Phys. Lett. B
401, 81 (1997).
\bibitem{51} V. Kher, N. Devlani, and A. K. Rai, Chin. Phys. C 41, 093101 (2017).
\bibitem{52} A.Upadhyay, M.Batra and P.Gupta, Prog. Theor. Exp. Phys. 53, 053B02 (2016).
\bibitem{53} Z. G. Wang, Phys. Rev. D 88, 114003 (2013).
\bibitem{54}  P. Gupta and A. Upadhyay, Phys. Rev. D 97, 014015 (2018).

\end{thebibliography}
\end{document}